\documentclass[twocolumn]{aastex631}
\usepackage{booktabs}
\usepackage{threeparttable}
\usepackage{bm}
\usepackage{comment}

\newcommand{\soa}{$\rm SO\ J=8_7 - 7_6\ $}
\newcommand{\sob}{$\rm SO\ J=5_6 - 4_5\ $}

\newcommand{\kms}{\rm km\ s^{-1}}
\newcommand{\Mjup}{M_{\rm Jup}}

\begin{document}

\title{Outflow Driven by a Protoplanet Embedded in the TW Hya Disk}

\author[0000-0001-8002-8473]{Tomohiro C. Yoshida}
\affiliation{National Astronomical Observatory of Japan, 2-21-1 Osawa, Mitaka, Tokyo 181-8588, Japan}
\affiliation{Department of Astronomical Science, The Graduate University for Advanced Studies, SOKENDAI, 2-21-1 Osawa, Mitaka, Tokyo 181-8588, Japan}

\author[0000-0002-7058-7682]{Hideko Nomura}
\affiliation{National Astronomical Observatory of Japan, 2-21-1 Osawa, Mitaka, Tokyo 181-8588, Japan}
\affiliation{Department of Astronomical Science, The Graduate University for Advanced Studies, SOKENDAI, 2-21-1 Osawa, Mitaka, Tokyo 181-8588, Japan}

\author[0000-0003-1413-1776]{Charles J. Law}
\altaffiliation{NASA Hubble Fellowship Program Sagan Fellow}
\affiliation{Department of Astronomy, University of Virginia, Charlottesville, VA 22904, USA}

\author[0000-0003-1534-5186]{Richard Teague}
\affiliation{Department of Earth, Atmospheric, and Planetary Sciences, Massachusetts Institute of Technology, Cambridge, MA 02139, USA}

\author[0000-0003-2993-5312]{Yuhito Shibaike}
\affiliation{National Astronomical Observatory of Japan, 2-21-1 Osawa, Mitaka, Tokyo 181-8588, Japan}

\author[0000-0002-2026-8157]{Kenji Furuya}
\affiliation{Department of Astronomy, The University of Tokyo, Bunkyo-ku, Tokyo 113-0033, Japan}

\author[0000-0002-6034-2892]{Takashi Tsukagoshi}
\affiliation{{  Faculty of Engineering, Ashikaga University, Ohmae-cho 268-1, Ashikaga, Tochigi, 326-8558, Japan}}



\begin{abstract}
Gas giant planets are formed by gas accretion onto planetary cores in protoplanetary disks. However, direct evidence of this process is still lacking, limiting our understanding of planetary formation processes.
During mass accretion, planet-driven outflows may be launched, which could be observable by shock tracers such as sulfur monoxide (SO).
We report the detection of SO gas in the protoplanetary disk around TW Hya in archival Atacama Large Millimeter/sub-millimeter Array (ALMA) observations.
The \soa emission line is detected at a $6\sigma$ significance and localized to the southeast region of the disk with an arc-like morphology.
The line center is red-shifted with respect to the systemic velocity by $\sim5\ \kms$.
The starting point of the SO emission is located at a planet-carved dust gap at $42$ au.
We attribute this to an outflow driven by an embedded protoplanet.
Indeed, the observed morphology is well reproduced by a ballistic outflow model.
The outflow velocity suggests that the outflow launching source has a mass of $\sim 4 M_\Earth$ and the mass-loss rate is $3\times10^{-8} - 1\times10^{-6}\ M_{\rm Jup}\ {\rm yr^{-1}}$.
With the relation of mass-loss and mass-accretion rates established for protostars, we estimated the mass-accretion rate onto the protoplanet to be $3\times10^{-7} - 1\times10^{-5}\ M_{\rm Jup}\ {\rm yr^{-1}}$, which matches theoretical predictions for a $\sim 4 M_\Earth$ planet at this separation.
The detection of planet-driven outflow provides us a unique opportunity to directly probe the earliest phase of gas giant planet formation.
\end{abstract}



\section{Introduction} \label{sec:intro}

Gas giant planets are formed by coagulation of dust and accretion of gas in protoplanetary disks according to the core accretion scenario.
It is predicted that gas accretion onto a protoplanet occurs through a circumplanetary disk \citep{canu06}.
However, the detailed process of gas accretion is still unclear.
To understand the gas accretion processes, it is crucial to reveal the physical conditions in the phase where protoplanets are still embedded in the protoplanetary disk.
The most reliable feature of gas accretion onto protoplanets is H$\alpha$ emission.
Indeed, for example, the confirmed protoplanets PDS 70 b and c exhibit this feature \citep{haff20} in addition to the detection of circumplanetary disks \citep{isel19, beni21}.
However, if the planet is deeply embedded in the natal protoplanetary disk, the H$\alpha$ line is hard to detect due to dust extinction \citep{marl22, alar24}.
Therefore, an alternative tracer of embedded protoplanets is needed.

Meanwhile, a similar issue occurs in the field of {\it star} formation.
Protostars do not show accretion features in H$\alpha$ emission since they are deeply embedded in the natal molecular cloud.
Alternatively, protostellar outflows have been broadly used to detect protostars \citep{arce07}.
Such outflows are often observed by shock tracers such as SO or SiO.
The outflows are launched following gas accretion to the central star due to the magneto-hydrodynamic effect.
This physical link is indeed confirmed by a tight correlation between the mass accretion rate and the outflowing mass loss rate \citep{ball16}.

Protoplanetary accretion would be a scaled-down version of protostellar accretion.
Therefore, as an analogy, we can expect that a similar process that is caused by the gas accretion --- {\it planet-driven outflow.}
The idea of planet-driven outflow was originally proposed by \citet{quil98} and further investigated by \citet{fend03}.
\citet{mach06} performed magneto-hydrodynamic (MHD) simulations and found that an outflow is driven from a protoplanet embedded in a disk with a velocity of ${\sim}10\ {\rm km\ s^{-1}}$.
\citet{gres13} predicted a mono-polar collimated outflow launched by a protoplanet using global MHD simulations of a protoplanetary disk.
{ Local but non-ideal MHD simulations by \citet{shib23} also show the possibility of a continuous outflow from a circumplanetary disk.}

Indeed, a few potential signatures of the planet-driven outflow have started to be found by ALMA observations.
\citet{alar22} identified a localized kinematic structure in [CI] emission in the HD 163296 disk, which can be interpreted as an outflow from a possible protoplanet.
\citet{law23} detected localized SO emission and potential SiS emission in the HD 169142 disk.
While the SO emission traces ice sublimation near the protoplanet HD 169142 b, the SiS emission might originate from a shock due to a potential outflow launched from the embedded planet.

In this Letter, we report the detection of SO gas that originates from a molecular outflow driven by an embedded protoplanet in the TW Hya disk.
The TW Hya disk is the closest protoplanetary disk from the Earth \citep[$D\sim 60$ pc; ][]{gaia16, gaia21} with a face-on geometry, and exhibits dust gaps which are much shallower than ones in the PDS 70 disk \citep{andr16, tsuk16}.
There are two major dust gaps at ${\sim} 26$ au and ${\sim} 42$ au from the central star, which could be carved by planets.
\citet{ment19} pointed out that super-Earths with a mass of $\sim4\ M_\Earth$ can explain the both gaps.
Furthermore, \citet{tsuk19} discovered a blob structure in its mm-dust disk.
This can be interpreted as a circumplanetary disk or dust-accumulated clump, and both of two scenarios imply ongoing planet formation.

The structure of this Letter is the following.
In the next section, we introduce the archival observations that we analyzed.
In Section \ref{sec:res}, we present the results, our modeling of the outflow, and the dynamical parameters of the outflow.
We then discuss the results in Section 4 and summarize this Letter in Section 5.

\section{Archival Observations} \label{sec:obs}

We analyzed two datasets in the ALMA science archive.
The \soa line at 340.714155 GHz (the upper state energy $E_{\rm up}\sim81 {\rm K}$) in Band 7 was observed on April 8 and May 21, 2017 with a baseline range of 15 - 1100 m (project ID: 2016.1.00311.S, PI: I.Cleeves).
These observations were performed as a part of the TW Hya Rosetta Stone project \citep{ober20}.
The total on-source integration time is $\sim 78$ min.
We obtained the visibility data from the archive and ran the pipeline calibration script. We then first performed iterative phase self-calibration with {\tt solints} $=$ 1,200 s, 400 s, 120 s, and 30 s, followed by amplitude self-calibration with an interval of the EB duration. This resulted in an improvement of the peak continuum SNR of a factor of $\sim5$.
Finally, self-calibration solutions were then applied to the line data.
We then imaged the \soa line and identified the emission.
We masked the emission manually (Appendix \ref{app:channel}) and CLEANed with a threshold of $2\sigma$, where $\sigma$ is the root mean square noise level.
We adopted natural weighting to achieve the highest point source sensitivity.
The channel width is set to $0.1\ \kms$, which is comparable to the native velocity resolution ($0.11\ \kms$).
The resultant beam size is $0\farcs48 \times 0\farcs30$ with a position angle of $2.6^\circ$. 
The noise level was $4.8\ {\rm mJy}\ {\rm beam^{-1}}$.

The \sob line at 219.949442 GHz ($E_{\rm up}\sim 35\ {\rm K}$) in Band 6 was observed on April 4 and 6, 2021 with the baseline range of 15 - 440 m (project ID: 2019.1.01177.S, PI: C. Eistrup) and total on-source time of $\sim 95$ min.
These observations were originally published by \citet{will23}.
The visibility data downloaded from the archive was analyzed in the same way as the \soa line; we performed iterative self-calibration with the same time intervals.
We made the image cube with the natural weighting and a channel width of $\sim 0.1\ \kms$, which is similar to the native velocity resolution ($0.10\ \kms$).
The resultant beam size and noise level are $0\farcs39 \times 0\farcs38, {\rm PA}=174.1^\circ$ and $2.3\ {\rm mJy}\ {\rm beam^{-1}}$.
Note that, for conservative purposes, we did not perform CLEAN for this line as we did not see a significant signal in each single channel.
The observation dates of the two SO lines are four years apart. In the following analysis, we corrected the proper motion of TW Hya by matching the peak of the continuum disk.

To supplement the SO lines, we also analyzed archival observations of the CO ${\rm J=3-2}$ line from \citet{yosh22} and imaged it with natural weighting.
We achieved a sensitivity of $1.3\ {\rm mJy\ beam^{-1}}$ at a 0.25 $\kms$ channel width and a beam size of $0\farcs27 \times 0\farcs21, {\rm PA} = -81.6^\circ$.

\section{Analysis and Results}
\label{sec:res}
\subsection{Localized SO emission}
\begin{figure}
\centering
\includegraphics[width=0.9\linewidth]{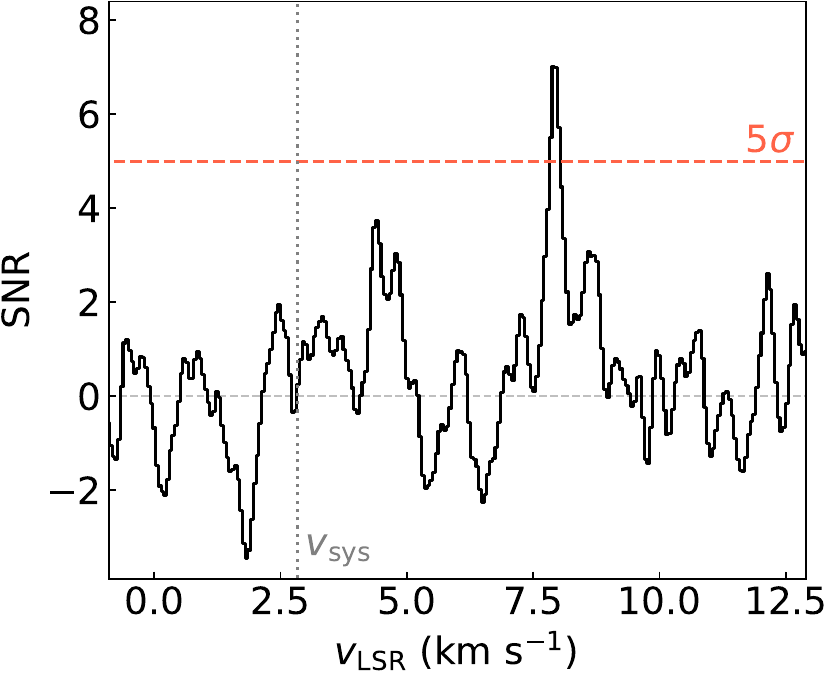}
\caption{Matched filter response of the \soa\ line with the CLEAN mask. The signal is detected with a $\sim~7\sigma$ significance at $v_{\rm LSR}\sim7.7\ \kms$.}
\label{fig:mf}
\end{figure}
We performed a matched filter analysis to confirm the detection of the \soa\ line using a python code {\tt VISIBLE}  \citep{loom18}.
With the mask used for CLEANing, we obtained the matched filter response spectrum as shown in Figure \ref{fig:mf}.
The signal is detected with a $\sim7\sigma$ significance at $v_{\rm LSR}\sim7.7\ \kms$.
We searched other molecular lines around this frequency but did not find any candidates.
We note that the systemic velocity $v_{\rm sys}$ is $\sim2.8\ \kms$ \citep{teag22} and the projected Keplerian rotation velocity is typically less than $\sim 1.0\ \kms$.
Therefore, we conclude that the emission likely traces SO gas with non-Keplerian vertical motion.
It is notable that we did not find any Keplerian component of SO. 

\begin{figure*}[htbp]
\centering
\includegraphics[width=0.9\linewidth]{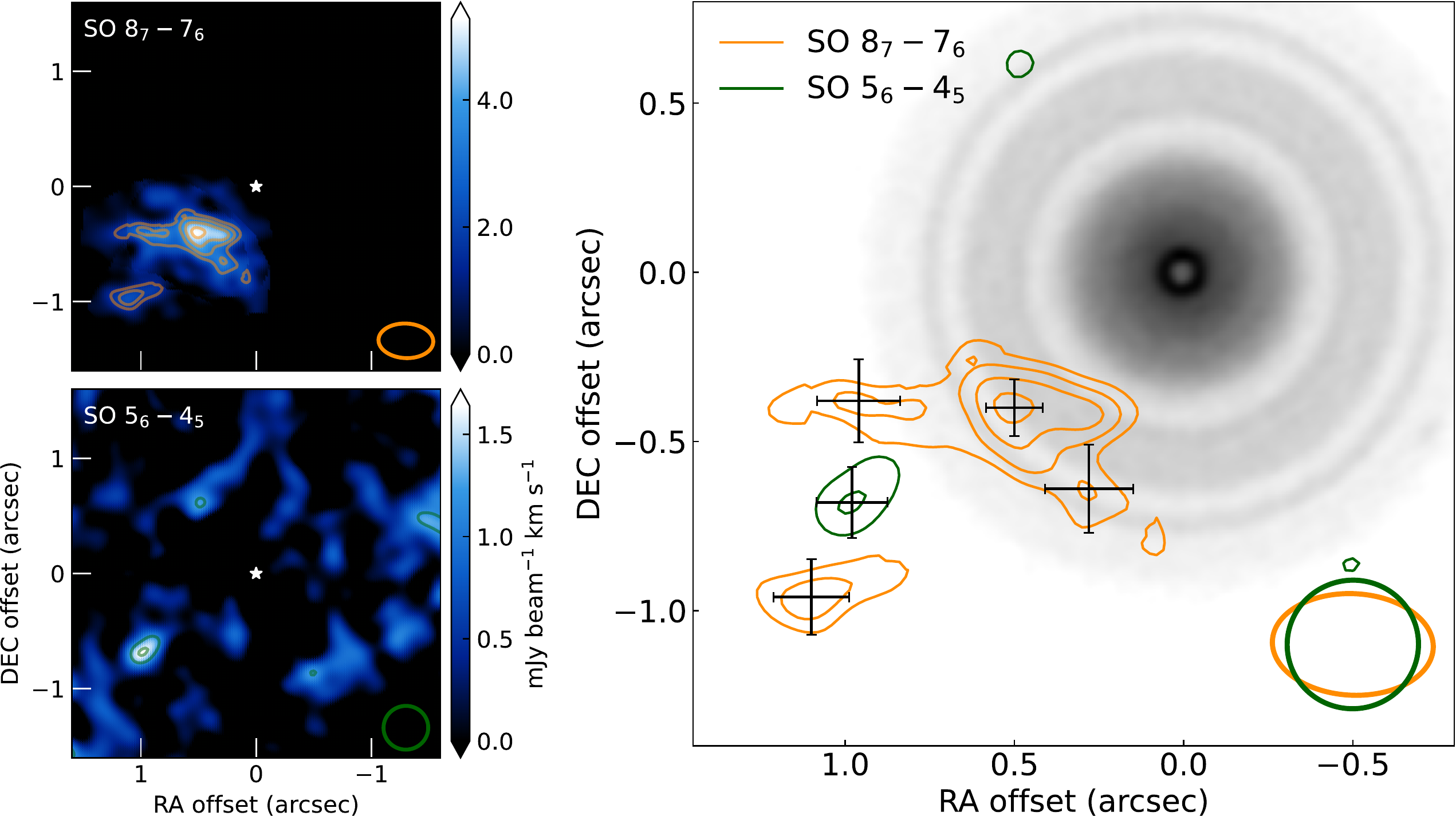}
\caption{(left) Integrated intensity maps of the \soa\ and \sob\ lines. The colored ellipses in the bottom right indicate the beam sizes. (right) SO emission overlaud in contours over the 233 GHz continuum image by \citet{tsuk19}. The color contours correspond to $[3,4,5,6]\sigma$. The black crosses show the peak position of knots with positional errors.}
\label{fig:mom0}
\end{figure*}
The channel map of the \soa\ line (Appendix \ref{app:channel}) also shows evident signals over three channels.
The emission is localized to the southeast region of the TW Hya disk.
We created integrated intensity maps of both SO lines (Figure \ref{fig:mom0}). Here, the CLEAN mask is used for the \soa\ line.
On the other hand, for the \sob\ line, we did not apply any mask as we did not perform CLEAN but just integrated over $v_{\rm LSR} = 7.6-7.8\ \kms$ where the \soa\ line is detected.
The \soa\ line emission is localized to the southeast part of the disk but has an extension over several beams.
The total integrated flux of the \soa\ line is measured to be $13\pm3\ {\rm mJy\ \kms}$.
In addition, we found evidence of the \sob\ line emission in a similar region.
For completeness, we also present the integrated intensity map without the CLEAN mask in Figure \ref{fig:mom0c2}.
In the right panel of Figure \ref{fig:mom0}, the signal-to-noise ratio (SNR) map of both lines is plotted in contours.
The peak of the \soa\ line reaches $\sim6\sigma$, confirming that the signal is real.
The \sob\ line peak is located at the region where the \soa\ line is not detected.
This is probably due to the difference in the excitation condition. Indeed, while the upper state energy of the \soa\ line is $\sim81$ K, that of the \sob\ line is just $\sim35$ K. The \soa\ line traces hotter gas than the \sob\ line.

In Figure \ref{fig:mom0}, we also plotted the 233 GHz high-resolution continuum image published by \citet{tsuk19}.
The \soa\ peak matches with the dust gap at $\sim42$ au from the central star \citep{andr16, tsuk16, maci21}, where the existence of a super-Earth mass planet is suggested by simulations \citep{ment19}.
The emission has an arc-like morphology that starts from the dust gap and continues to the outer region.

\begin{figure}[htbp]
\centering
\includegraphics[width=0.9\linewidth]{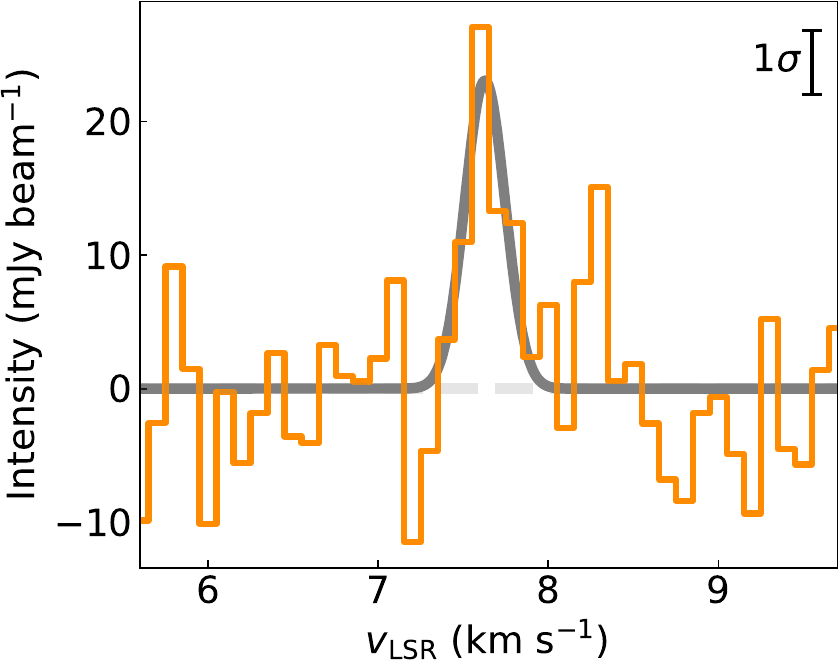}
\caption{ \soa spectrum at the emission peak. The best-fitted Gaussian is also described in the gray-thick line. }
\label{fig:peak}
\end{figure}
We extracted the \soa\ spectrum at the peak position and plotted it in Figure \ref{fig:peak}.
The peak signal-to-noise ratio in this spectrum reaches 5.6$\sigma$.
To measure the line width, we fit a Gaussian to the spectrum and obtained an FWHM of $0.28\ \kms$, which means that the line is spectrally resolved as the velocity resolution is $\sim0.1\ \kms$.

\subsection{Outflow trajectory model}
\label{sec:model}

In protostellar and protoplanetary disks, SO emission is mainly detected in thermally heated or dynamically shocked regions.
The accretion shock to protostellar disks \citep[e.g.,][]{saka14} is one of the typical examples.
In Class II disks, \citet{boot23} and \citet{law23} found SO emission that follow Keplerian rotation and attributed them to locally heated gas around protoplanets in two separate disks.
In the context of star formation, SO emission is frequently detected in protostellar outflows launched from the central protostar \citep[e.g.,][]{podi21}.
The SO emission in the TW Hya disk is spatially localized with a significant velocity deviation from the Keplerian rotation and has a peak at the dust gap.
This is similar to SiS emission in the HD 169142 disk \citep{law23}.
Moreover, the emission morphology resembles protostellar outflows \citep[e.g.,][]{ball16} except for the arc-like shape; there are several peaks similar to the ``knots'' of protostellar outflows although we note that this could be due to the low signal-to-noise ratio.
We propose that the SO emission traces an outflow from an embedded protoplanet in the dust gap, in other words, planet-driven outflow.

An intriguing feature of the SO emission is its arc-like morphology, which is not similar to protostellar outflows that are generally linearly well-aligned \citep[e.g.,][]{podi21}.
One of the major differences between the protostellar and planet-driven outflows is the motion of the outflow driving source.
In the protostellar case, the driving source (protostar) and its outflow follow inertial motion in interstellar space, which is usually not significant against the outflow motion.
On the other hand, in the protoplanetary case, the driving source (protoplanet) and its outflow feel gravity force from the central star, which may change the outflow morphology.

To test this case, we constructed a simple ballistic trajectory model.
First, we assume that an outflow knot follows the equation of motion
\begin{equation}
\label{eq:eom}
    \frac{d^2{\bm r}}{dt^2} = -\frac{GM_\star}{|\bm{r}|^3} \bm{r},
\end{equation}
where $\bm r$ is the positional vector, $G$ is the gravitational constant, and $M_\star$ is the stellar mass of TW Hya.
Here we only consider the gravity force from the central star since the protoplanet mass would be negligible in the observed outflow.
In addition to gravity force, other forces such as the pressure gradient force and magnetic force could work.
Therefore, with $z$-axis being the disk's vertical direction, we modified the $z$ component of Eq.\ref{eq:eom} to
\begin{equation}
   \frac{d^2{z}}{dt^2} = -\frac{GM_\star}{|\bm{r}|^3} z ( 1 - \epsilon ),
\end{equation}
where $\epsilon$ controls the vertical force balance.
For instance, if the knot is in a disk with hydrostatic equilibrium, $\epsilon$ should be unity and the vertical net force becomes zero.
On the other hand, if the knot is being accelerated by e.g. magnetic fields, $\epsilon > 1$. 
Then, we assume that the outflow knots are launched at an initial velocity of ${\bm v}$ from the protoplanetary system.
The protoplanet is assumed to be located at the dust gap radius of 42 au from the central star with an azimuthal angle of $\theta_p$ from the disk major axis.
With ${\bm v_p}$ being the orbital velocity of the protoplanet, the outflow has a velocity of ${\bm v} + {\bm v_p}$ with respect to the central star.
The outflow launching axis (i.e. the direction of ${\bm v}$) has an inclination angle from the normal of the disk $\phi$ and a position angle of $\psi$ from the major axis of the disk.
By solving Equation \ref{eq:eom} with those initial conditions, and projecting to the plane of the sky, we can calculate the trajectory of knots in the RA-DEC-$v_{\rm LOS}$ space.

We specified the knot position and line of sight velocity by taking the five emission peaks in the image cubes.
We estimated the errors in RA and DEC using the nominal astrometric accuracy of ALMA (ALMA Technical Handbook). For the velocity, we conservatively assume an uncertainty of twice the velocity resolution, $0.2\ \kms$.
Then, we set $v_{\rm out} (\equiv |{\bm v}|)$, $\phi$, $\psi$, $\epsilon$, $\theta_p$, and the dynamical time scale of the outflow $t_{\rm dyn}$ as free parameters and produce the outflow trajectory.
The trajectory is compared with the observed knot positions and velocity through a log-likelihood function,
\begin{equation}
    \log P = - \sum_i \sum_d \frac{D_{i, d}^2}{2 \sigma_{i, d}^2}.
\end{equation}
Here $i\ (1 \leq i \leq 5)$ and $d$ indicate indices of a knot and one of the dimensions (RA, DEC, and LOS velocity), respectively.
$D_{i, d}$ is the orthogonal distance between the knot $i$ and the trajectory along the $d$ axis.
$\sigma_{i, d}$ is the standard deviation of the position and velocity.

We sampled the posterior probability distribution using the { Markov}-Chain Monte-Carlo (MCMC) method with a python module {\tt emcee} \citep{fore13}.
While we applied uniform priors for most parameters, we adopted a one-sided Gaussian with a mean of $185^\circ$ and a standard deviation of $30^\circ$ for $\theta_p$ so that the protoplanet is located near the first ({ rightmost}) knot but does not exceed it assuming clock-wise rotation.
We employed 24 walkers with 5000 steps (and 2000 burn-in steps) to sample the posterior distribution.
The MCMC sampling was well-converged and the corner plot is shown in Appendix \ref{app:corner}.
As a result, the median values with 16-84th percentiles of the marginal posterior distribution are found to be $v_{\rm out} = 5.6^{+0.5}_{-0.4}\ \kms$, $\phi = -133{^\circ}^{+5}_{-5}$, $\psi = -80{^\circ}^{+7}_{-7}$, and $\epsilon = 1.15^{+0.04}_{-0.03}$, $185{^\circ} < \theta_p \lesssim 210{^\circ}$, and $t_{\rm dyn} \gtrsim 100\ {\rm yr}$, respectively.
\begin{figure*}[htbp]
\centering
\includegraphics[width=0.9\linewidth]{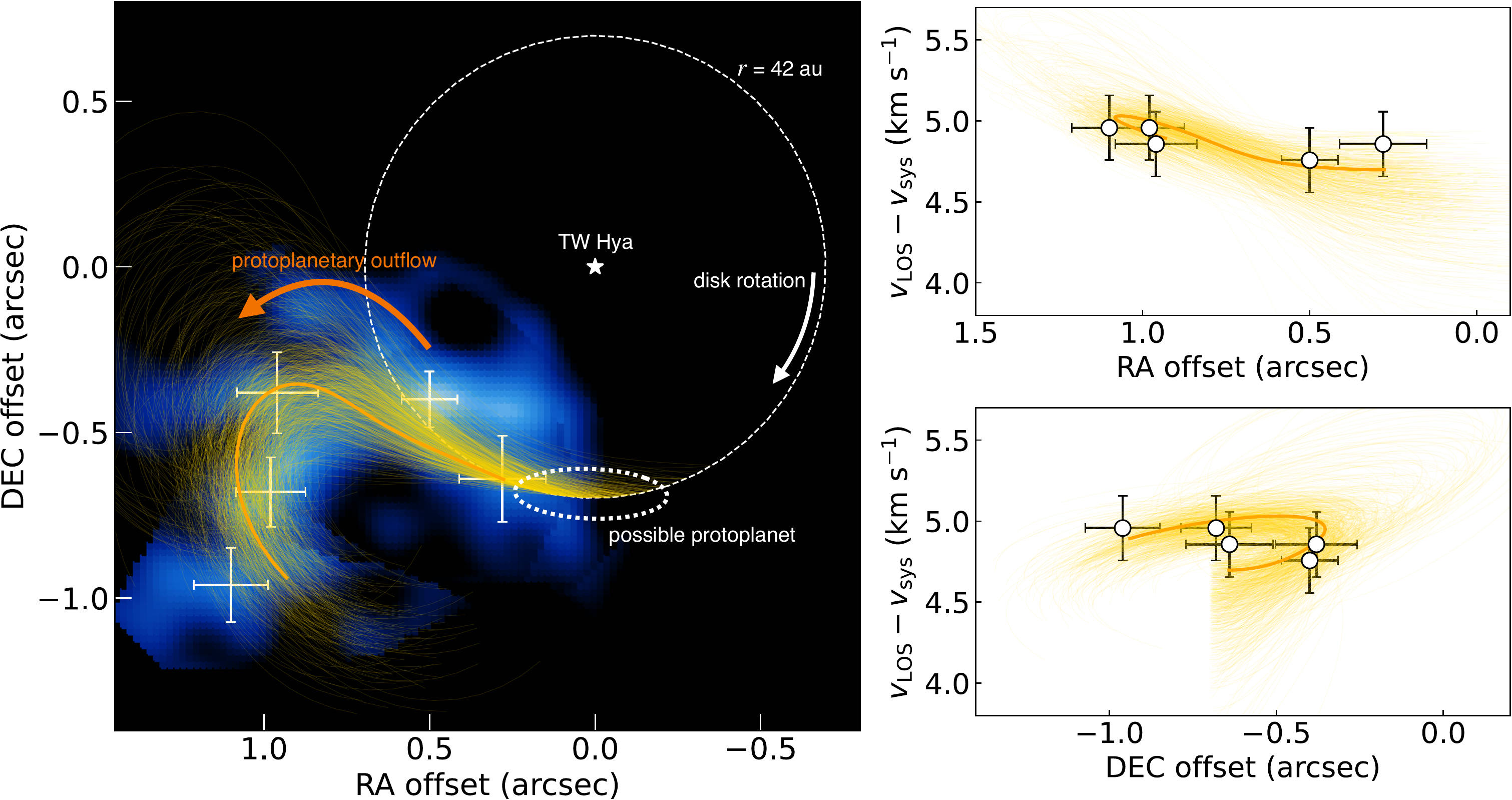}
\caption{(left) Best-fit outflow trajectory is plotted in the orange thick line. The trajectories calculated from the parameters sampled by the posterior distribution are plotted in yellow thin lines. The background color map shows the sum of the two signal-to-noise ratio integrated intensity maps. The white crosses indicate the knot location used for fitting. 
(right) Outflow trajectories in the spatial-velocity domains.
}
\label{fig:mom02}
\end{figure*}
The best-fit trajectory is well-fitted to the observed emission morphology as shown in Figure \ref{fig:mom02}.

In summary, the observed morphology of the SO emission can be well explained by the simple ballistic outflow model.
In this scenario, the outflow is launched from a circumplanetary disk around an embedded protoplanet that is located in the southern part of the 42 au dust gap.
The outflow axis is likely tilted from the normal of the protoplanetary disk by $\sim50^\circ$.
The most distant SO knot was launched $\sim100\ {\rm yr}$ ago. The knots would be feeling small acceleration in the vertical direction.
\begin{figure}[htbp]
\centering
\includegraphics[width=0.9\linewidth]{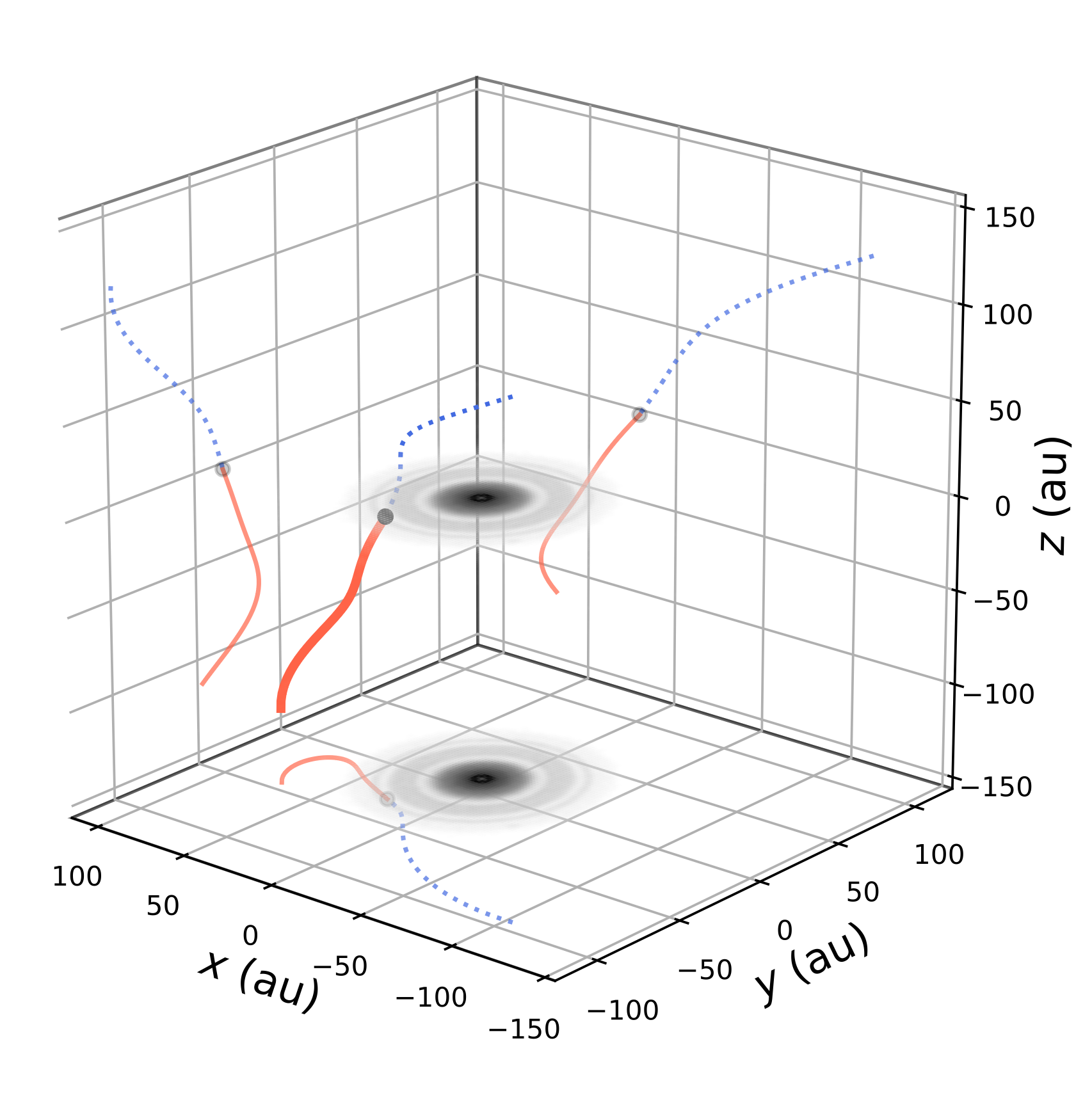}
\caption{Best-fit outflow trajectory in the three-dimensional disk coordinate is plotted in the thick red line. The dotted blue line indicates the blue-shifted outflow if it exists.}
\label{fig:3D}
\end{figure}
We also show the 3D morphology of the best-fitted model in Figure \ref{fig:3D}.
In the vertical direction, the outflow should reach $\sim100$ au from the disk plane.
In Figure \ref{fig:trajects}, the time evolution of the trajectory from 80 yr ago until now is plotted, highlighting the qualitative difference in morphology of protostellar and planet-driven outflows.

\subsection{Dynamical Parameters}
\label{sec:dynamicparams}
Assuming that the emission is optically thin and local thermal equilibrium is valid, we estimated the total mass of SO molecules to be $(1-4) \times 10^{20}\ {\rm g}$ and the averaged SO column density of $(1-4) \times 10^{12}\ {\rm cm^{-2}}$ for a temperature range of $30-300$ K from the integrated flux of the \soa\ line ($13\pm3\ {\rm mJy\ \kms}$).
To convert the SO mass to the total gas mass of the outflow $M_{\rm out}$, it is necessary to assume the SO/H$_2$ ratio in the outflow.
As a rough upper limit of SO/H$_2$, we can consider a half of the cosmic S/H ratio, $\sim6.6\times 10^{-6}$ \citep{aspl21}.
On the other hand, it is known that SO is significantly depleted in dense molecular clouds.
We could take the SO/H$_2$ ratio in molecular clouds, $\sim10^{-8}$ \citep{ruff99}, as a rough lower limit.
In the case of protostellar outflows, \citet{podi21} estimated the SO/H$_2$ ratio of $10^{-7}-10^{-6}$ in detected SO outflows.
In a protoplanetary disk around HD 100546, where thermally desorbed SO is detected, SO/H$_2$ is constrained to be $3.5\times10^{-7} - 5.0\times10^{-6}$ depending on chemical models \citep{boot18, boot23}. 
Overall, it would be reasonable to assume the SO/H$_2$ ratio in the outflow lies in $10^{-7} - 10^{-6}$.
With this bound of SO/H$_2$, we can estimate $M_{\rm out}$ to be $3\times10^{-6}-1\times10^{-4}\ \Mjup$.

We searched for a counterpart of the SO emission in the deep CO $J=3-2$ line observations (Section \ref{sec:obs}), however, we did not find any signal in the corresponding spatial and velocity range at $v_{\rm LSR} \sim 7.7\ \kms$.
For the same area with the \soa\ mask, we put a $3\sigma$ upper limit of the integrated CO ${\rm J=3-2}$ line flux of $5.4\ {\rm mJy\ \kms}$.
Using this value, we estimate the upper limit of $M_{\rm out}$ to be $5\times10^{-7} - 2\times10^{-4}\ \Mjup$ for the typical CO/H$_2$ ratios of $10^{-6}-10^{-4}$ \citep[e.g.,][]{zhan19} and the temperature range of $30-300$ K.
This is consistent with the mass obtained by SO.

The trajectory modeling suggests that the dynamical time scale, or the time since ejection of the most distant knot, is $\sim100$ yr.
With this value, the mass-lass rate of the outflow $\dot{M}_{\rm out} \equiv M_{\rm out}/t_{\rm dyn}$ is estimated to be $3\times10^{-8}-1\times10^{-6}\ \Mjup\ {\rm yr}^{-1}$.
In addition, we estimated the momentum $M_{\rm out} v_{\rm out}$ and momentum supply rate $\dot{M}_{\rm out} v_{\rm out}$.
\begin{table*}[hbtp]
  \caption{Dynamical Parameters}
  \label{tab:params}
  \centering
  \begin{tabular}{lccr}
    \hline \hline
    Parameters & Symbols & Values & Units \\
    \hline \hline
    Velocity & $v_{\rm out}$ & $5.6^{+0.5}_{-0.4}$ & $\kms$ \\
    Dynamical timescale & $t_{\rm dyn}$ & $\sim100$ & ${\rm yr}$ \\
    Mass & $M_{\rm out}$ & $3\times10^{-6}-1\times10^{-4}$ & $\Mjup$ \\
    Mass-loss rate & $\dot{M}_{\rm out}$ & $3\times10^{-8}-1\times10^{-6}$ & $\Mjup\ {\rm yr}^{-1}$ \\
    Momentum & $M_{\rm out} v_{\rm out}$ & $2\times10^{-5} - 6\times10^{-4}$ & $\Mjup\ \kms$ \\
    Momentum supply rate & $\dot{M}_{\rm out} v_{\rm out}$ & $2\times10^{-7} - 6\times10^{-6}$ & $\Mjup\ \kms {\rm yr}^{-1}$ \\
    \hline
  \end{tabular}
\end{table*}
The derived dynamical parameters are summarized in Table \ref{tab:params}.

\section{Discussion} \label{sec:sum}

\subsection{Outflow-driving source}
The detection of the planet-driven outflow suggests the existence of a circumplanetary disk around the embedded protoplanet.
Major mechanisms that drive protostellar outflows such as the magneto-centrifugal disk wind \citep{pell92} and X-wind \citep{shu94} predict that the terminal speed of outflow is roughly
\begin{equation}
    v_{\rm out} \sim 3 v_c,
\end{equation}
where $v_c$ is the rotation velocity at the outflow launching point \citep{quil98}.
We apply this to the observed planet-driven outflow launched from the circumplanetary disk.

The rotation velocity in the circumplanetary disk around the outflow-driving protoplanet can be written as 
\begin{equation}
    v_c = \sqrt{\frac{G M_p}{R_l}},
\end{equation}
assuming Keplerian rotation.
Here, $M_p$ and $R_l$ are the protoplanet's mass and radius of the launching point measured from the protoplanet.
Therefore, the protoplanet's mass can be expressed as
\begin{equation}
    M_p \sim 6 \times 10^{-5} \left( \frac{R_l}{R_{\rm Jup}} \right) \left( \frac{v_{\rm out}}{1\ \kms} \right)^2\ \Mjup,
\end{equation}
with $R_{\rm Jup}$ being the Jupiter radius.
It is challenging to estimate $R_l$, however, we roughly assume $R_l\sim6R_{\rm Jup}$ by following \citet{quil98}, which is motivated by the orbital radius of the innermost Jovian satellite Io.
By substituting $v_{\rm out}\sim5.6\ \kms$, we obtain { $M_p \sim 4\ M_\Earth$ ($0.01\ \Mjup$).}

A conventional way to estimate a planet's mass is to focus on dust gap structures.
For the 42 au gap in the TW Hya disk, \citet{ment19} performed 
global three-dimensional hydrodynamics simulations and found that a $\sim4\ M_\Earth$ planet can explain the gap depth.
Our estimate from the outflow matches their estimate.

It is not trivial that a circumplanetary disk can be formed around a super-Earth.
For example, \citet{fung19} predicted that a $3.5\ M_\oplus$ planet can possess a circumplanetary disk in the case of efficient cooling, while \citet{krap24} suggested that a circumplanetary disk cannot be formed even around a Jupiter mass planet under some conditions.
Our results imply that a $4\ M_\oplus $ planet at 42 au can have a circumplanetary disk, constraining disk formation processes.

\citet{teag22} investigated velocity deviations in the CO and CS lines from the Keplerian rotation in the TW Hya disk. In both lines, they show $\sim 5 \sigma$ signals at the protoplanet position inferred by our analysis although it is unclear if a super-Earth mass planet can produce such kinematic signatures.

\subsection{Mass-accretion rate and luminosity}
There is a linear relation between the mass-accretion rate from circumstellar disks to protostars and the mass-loss rate by protostellar outflows, namely, $\dot{M}_{\rm out} \sim 0.1 \dot{M}_{\rm acc}$, where $\dot{M}_{\rm acc}$ is the mass-accretion rate of a protostar \citep[e.g.,][]{ball16}.
Assuming that this relation is valid in the case of planet-driven outflows, the mass-accretion rate of the protoplanet $\dot{M}_{p, \rm acc}$ is found to be $3\times10^{-7}-1\times10^{-5}\ \Mjup\ {\rm yr}^{-1}$.

Theoretical studies predict the mass-accretion rate as a function of protoplanetary disk parameters.
For instance, \citet{tani16} derived 
\begin{equation}
    \dot{M}_{p, \rm acc} = 0.29 \left( \frac{h_p}{r_p} \right)^{-2} \left( \frac{M_p}{M_\star} \right)^{4/3} \Sigma_g r_p^2 \Omega_p, 
\end{equation}
where $r_p, h_p, \Sigma_g, $ and $\Omega_p$ are the distance of a protoplanet from the central star, scale height of the protoplanetary disk at $r_p$, gas surface density of the protoplanetary disk, and Keplerian angular frequency at $r_p$, respectively.
Using the measured midplane gas temperature of $15$ K and $\Sigma_g$ of $10\ {\rm g\ cm^{-2}}$ at a radius of 42 au \citep{cala21}, we obtain $\dot{M}_{\rm acc} = 1.5 \times 10^{-6}\ \Mjup\ {\rm yr}^{-1}$, which is consistent with our estimation.
Note that this value does not depend on disk turbulence.

The accretion luminosity of the protoplanet is expressed as
\begin{equation}
    L = \frac{ G M_p \dot{M}_{p, \rm acc} }{R_p},
\end{equation}
with $R_p$ being the protoplanet's radius.
Assuming that the protoplanet density is the density of water ice, $\sim 4\ {\rm g\ cm^{-3}}$ \citep{liss14}, since it may consist of icy material, we can estimate $R_p \sim 1\times10^{9}\ {\rm cm}$.
Substituting all values, we obtain $L = 6\times10^{-6} - 2\times10^{-4}\ L_\odot$.
In protostellar outflows, it is known that the momentum supply rates are correlated with the bolometric luminosity of protostars. If we extrapolate the relation in \citet{ball16}, we can estimate the momentum supply rate of the outflow to be 
$(1-10)\times 10^{-5}\ \Mjup\ \kms {\rm yr}^{-1}$, which is 20-70 times larger than our estimate from the observed value ($2\times10^{-7} - 6\times10^{-6}\ \Mjup\ {\rm yr}^{-1}\ \kms$; Section \ref{sec:dynamicparams}) but within the scatter of protostellar cases.
This may suggest that gas accretion onto a protoplanet occurs by similar processes to protostellar accretion.

\subsection{Tilted outflow axis}
{ 
The tilt angle of the outflow launching axis from the protoplanetary disk plane is found to be $\sim50^\circ$.
The physical origin of this tilt is unclear.
\citet{gres13} found a time variability of the tilt angle of a circumplanetary disk due to stochastic accretion, which could result in the tilt of the outflow, although the degree of the tilt is only $\sim10^\circ$.
It is also known that the spin axes of some solar system planets are inclined with respect to the equatorial plane.
One of the explanations for this tilt is a spin-orbit resonance when the protoplanet is surrounded by its circumplanetary disk \citep{rogo20}.
This process could also tilt the outflow axis.
}

\subsection{Consistency with non-detections}
We already mentioned that the non-detection of the CO $J=3-2$ line is consistent with the SO detection (Section \ref{sec:dynamicparams}). Other transitions with higher upper-state energy would be useful to robustly determine the molecular abundances in the outflow.

The outflow trajectory model suggests that the blue-shifted outflow would be seen at $\sim -4.2\ \kms$ with respect to the systemic velocity if it exists (see also Figure \ref{fig:3D}, here we assumed that the blue-shifted outflow is launched in the opposite direction to the red-shifted side.).
We checked the image cube but did not see any signature.
Therefore, the planet-driven outflow may be mono-polar.
Indeed, an outflow predicted in the three-dimensional simulation by \citet{gres13} was mono-polar.
The SiS emission in the HD 169142 disk is detected only in the blue-shifted velocity range \citep{law23}.
It would be implied that mono-polar morphology is common (at least not rare) among planet-driven outflows.
We note that some protostellar outflows are also mono-polar \citep[e.g.,][]{wu04, louv18}.

The SO emission is not detected beyond $\sim 1''$ from the central star.
The chemical depletion timescale of SO should be much longer than the dynamical timescale of the outflow ($\sim100$ yr).
One possibility is that the outflow has been active only for this $\sim100$ yr. In this case, it implies that the protoplanetary accretion is periodic or stochastic, which is proposed as a solution for the low detection rate of protoplanets \citep{brit20}.
Another explanation is efficient cooling of the outflow in the downstream \citep[e.g.,][]{yosh21}.

\citet{boek17} reported a non-detection of any point sources in the TW Hya disk with VLT/SPHERE H-band observations and gave a detection limit of $\sim14$ magnitudes.
Assuming that the bolometric luminosity of the protoplanet is dominated by the accretion luminosity, we estimate the blackbody temperature of $2000-5000$ K with the planetary radius of $R_p\sim10^9\ {\rm cm}$ and calculate the flux.
Then, we estimate the contrast of the protoplanet to the central star (with a photospheric temperature of $4000$ K with a radius of $2R_\odot$) to be $10-13$ magnitudes from optical to infrared wavelength.
This contradiction may be solved by considering the extinction due to the small dust grains.
Indeed, the total extinction at H-band was estimated to be $4-5$ magnitudes in gaps seen at 21 and 85 au \citep{boek17}.
Since the mm-gaps at 42 au correspond to a ring in the H-band, the extinction at the H-band should be more severe.
Therefore, the actual contrast should be substantially larger than $14$ magnitudes.
Future deeper near-infrared observations could detect the protoplanet.

In addition, in the $42$ au gap, no circumplanetary disk candidate has been detected in the mm-wavelength so far despite deep observations \citep{tsuk19, maci21}.
The typical upper limit of the point source flux in their Band 6 or 7 images is given as $\sim 50-100\ {\rm \mu Jy}$.
Recent modeling study by \citet{shib24} pointed out that $M_p \dot{M}_{p, \rm acc} \gtrsim 10^{-6}\ {M^2_{\rm Jup} } {\rm yr^{-1}} $ is needed for a $\sim 100\ {\rm \mu\ Jy}$ emission although they assumed the PDS 70 system.
Our results of $M \dot{M}_{\rm acc} \sim 4\times10^{-9} - 1\times10^{-7} \ {M^2_{\rm Jup} } {\rm yr^{-1}} $ is consistent with the non-detection of a circumplanetary disk.
We note that the mass and mass accretion rates are also consistent with $\rm L'$ band observations by \citet{ruan17}.

\subsection{Open Questions}
\label{sec:oq}

Several open questions are now arising.
The line width of \soa\ is $\sim0.28\ \kms$, which is almost consistent with a thermal line width (c.f., thermal line width at 100 K is $\sim0.3\ \kms$).
On the other hand, most protostellar outflows exhibit line widths of a few to $\sim10\ \kms$ \citep{podi15}.
An exceptionally narrow linewidth was observed in the SiO line toward L1448-mm by \citet{jime04}.
\citet{jime09} successfully reproduces the line profile by the magnetic precursor of C-shocks. 
It is unclear if the SO emission in the TW Hya disk is related to this or not. Future modeling efforts are required.

\citet{quil98} indicated that a magnetic field strength of $\sim10$ G in a circumplanetary disk is required to launch planet-driven outflows. However, it would not be trivial that the protoplanet possesses such strong fields although \citet{fend03} suggested that it is possible.

Finally, we note that follow-up observations will explore if the outflow can be observed in other molecules such as SiS \citep{law23}, providing fruitful information on the physical and chemical conditions.

\section{Summary}

We detected asymmetric emission from the \soa\ and \sob\ lines in the nearest protoplanetary disk around TW Hya.
The emission significantly deviates from the systemic velocity of TW Hya, is localized in the southeast part of the disk, and peaks at a radius of 42 au from the central star where the existence of a super-Earth mass planet has been suggested.
The emission morphology resembles protostellar outflows but is arc-like.
We successfully reproduced the morphology in the spatial and velocity space with a simple ballistic outflow model and concluded that the SO traces the planet-driven outflow launched by an accreting protoplanet at a radius of 42 au.

With the outflow velocity derived by the modeling, we estimated the protoplanet mass to be $\sim4\ M_\Earth$ assuming that the velocity is determined by the gravitational potential, which is in excellent agreement with previous results based on the dust gap \citep{ment19}.
Assuming conversion factors, we also derived the mass-loss rate of the outflow.
Using the well-known relation between the mass-loss rate and mass-accretion rate, which is obtained by protostellar outflow studies, we constrained the mass-accretion rate of $3\times10^{-7}-1\times10^{-5}\ \Mjup\ {\rm yr}^{-1}$.
These values are consistent with theoretical studies.
The embedded super-Earth mass planet at the 42 au gap is in the main gas accretion phase.
{ It is also suggested that the outflow launching axis is tilted by $\sim50^\circ$, although its physical origin is unclear.
}

The detection of outflow driven by an embedded planet gave us a unique opportunity to test { 
the theory of the earliest phase of the gas giant planet formation} for the first time. 
Future observations of shock tracers including SO and other higher $E_{\rm up}$ lines in TW Hya and other disks will significantly improve our understanding.

\begin{acknowledgments}
This Letter makes use of the following ALMA data:
ADS/JAO.ALMA\#2016.1.00311.S, 2019.1.01177.S, 2015.1.00686.S, 2016.1.00629.S, 2018.1.00980.S, 2016.1.00842.S, and 2017.1.00520.S.
ALMA is a partnership of ESO (representing its member states), NSF (USA) and NINS (Japan), together with NRC (Canada), NSTC and ASIAA (Taiwan), and KASI (Republic of Korea), in cooperation with the Republic of Chile. The Joint ALMA Observatory is operated by ESO, AUI/NRAO and NAOJ.
This work was supported by Grant-in-Aid for JSPS Fellows, JP23KJ1008 (T.C.Y.).
Support for C.J.L. was provided by NASA through the NASA Hubble Fellowship grant No. HST-HF2-51535.001-A awarded by the Space Telescope Science Institute, which is operated by the Association of Universities for Research in Astronomy, Inc., for NASA, under contract NAS5-26555.
\end{acknowledgments}

%

\vspace{5mm}
\facilities{ALMA}


\software{astropy \citep{astropy:2013, astropy:2018, astropy:2022},  CASA \citep{mcmu07}, emcee \citep{fore13} }



\appendix

\section{Channel maps and the integrated intensity map without masking}
\label{app:channel}
\begin{figure*}[htbp]
\centering
\includegraphics[width=1.0\linewidth]{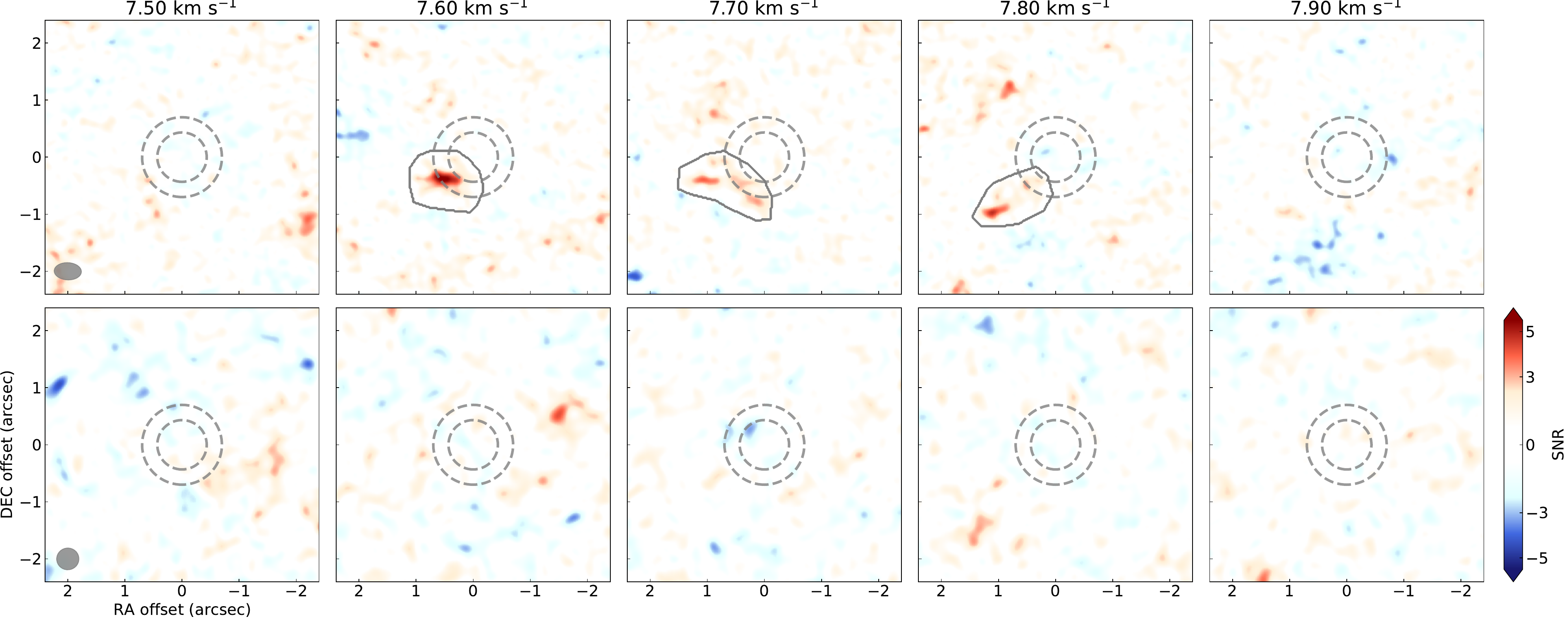}
\caption{Channel maps of the \soa\ line (upper row) and \sob\ line (lower row). We normalized the intensity by the noise level. The { grey} dotted circles indicate the two major dust gaps at 26 and 42 au from the central star. The { grey} solid lines mark the hand-written CLEAN masks. The gray ellipses in the bottom left corner show the synthetic beam sizes.}
\label{fig:channelmap}
\end{figure*}
Channel maps of the \soa\ and \sob\ lines are displayed in Figure \ref{fig:channelmap}.

\begin{figure}[htbp]
\centering
\includegraphics[width=0.4\linewidth]{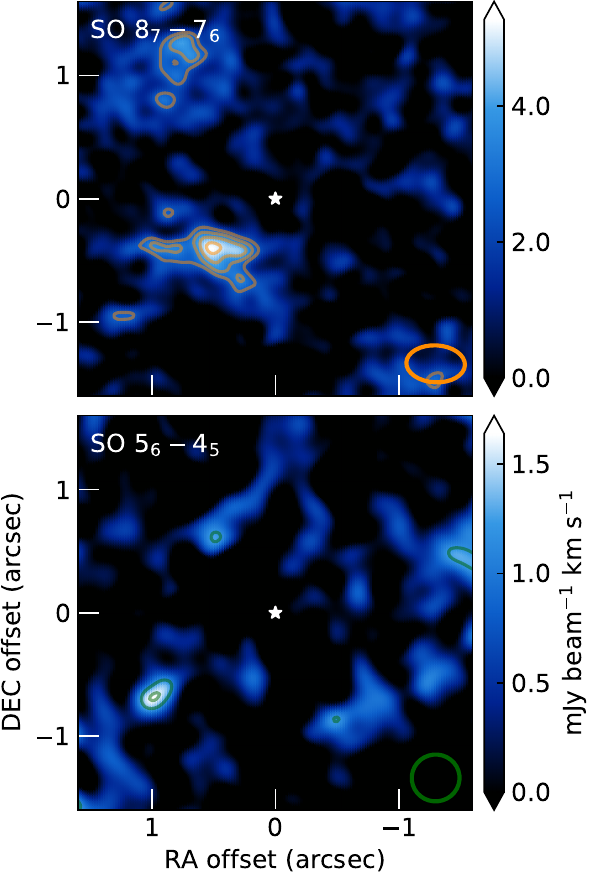}
\caption{Same as the left panel of Figure \ref{fig:mom0}, but without applying a CLEAN mask.}
\label{fig:mom0c2}
\end{figure}

We also plotted the integrated intensity map of the \soa\ line without CLEAN mask in Figure \ref{fig:mom0c2}. We just integrated the channel map with a velocity range of $7.6-7.8\ \kms$ where the \soa\ emission is detected.

\section{Corner plot of the MCMC fitting and time evolution of the outflow trajectory}
\label{app:corner}
\begin{figure*}[htbp]
\centering
\includegraphics[width=1.0\linewidth]{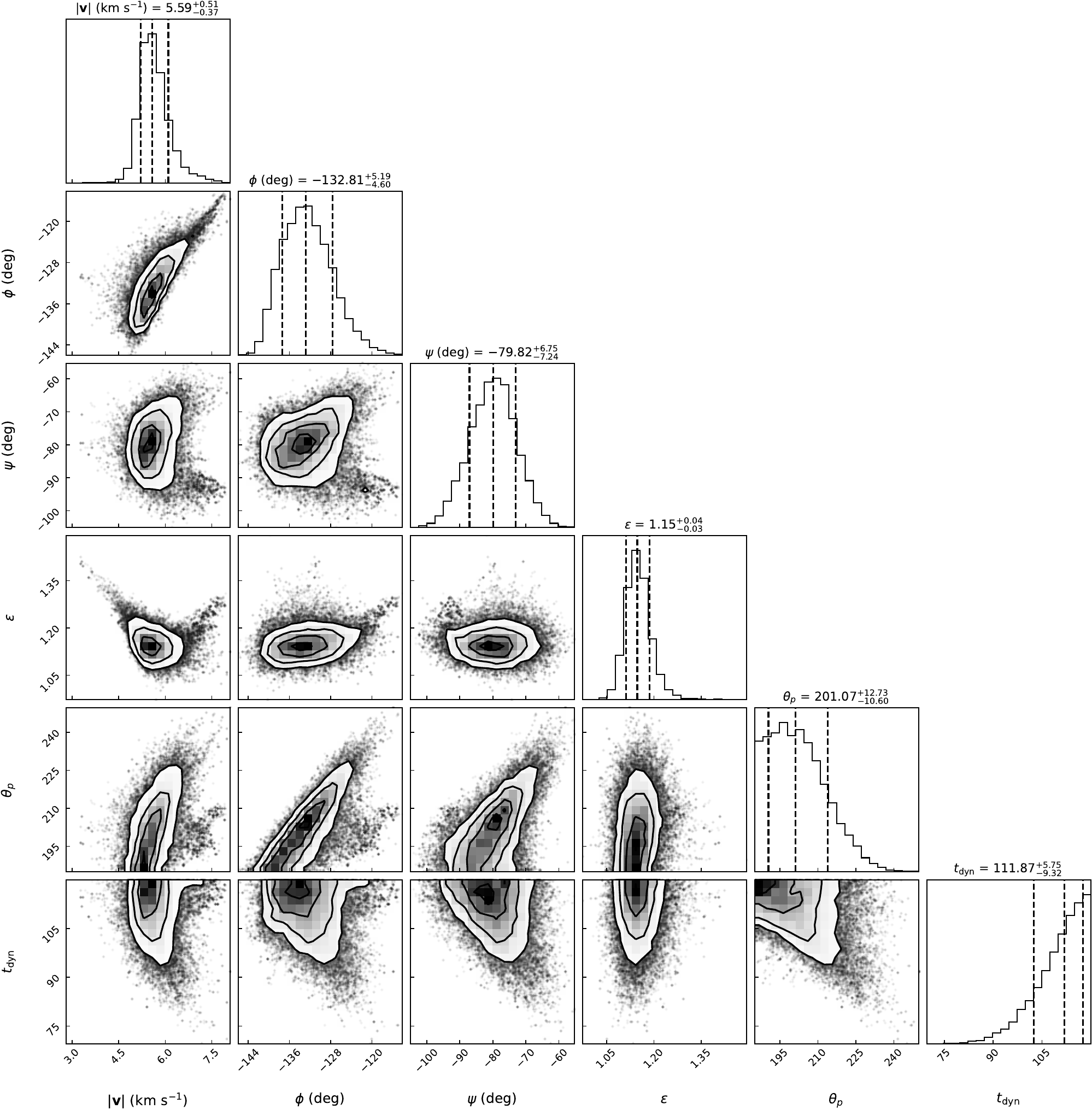}
\caption{Corner plot of our trajectory model fitting. Note that the values and errors in the panel title indicate the median and $[16, 84] \%$ percentiles.}
\label{fig:corner}
\end{figure*}
The marginal probability distributions of each parameter in our trajectory model fitting (Section \ref{sec:model}) are shown in Figure \ref{fig:corner}.

\begin{figure*}[htbp]
\centering
\includegraphics[width=1.0\linewidth]{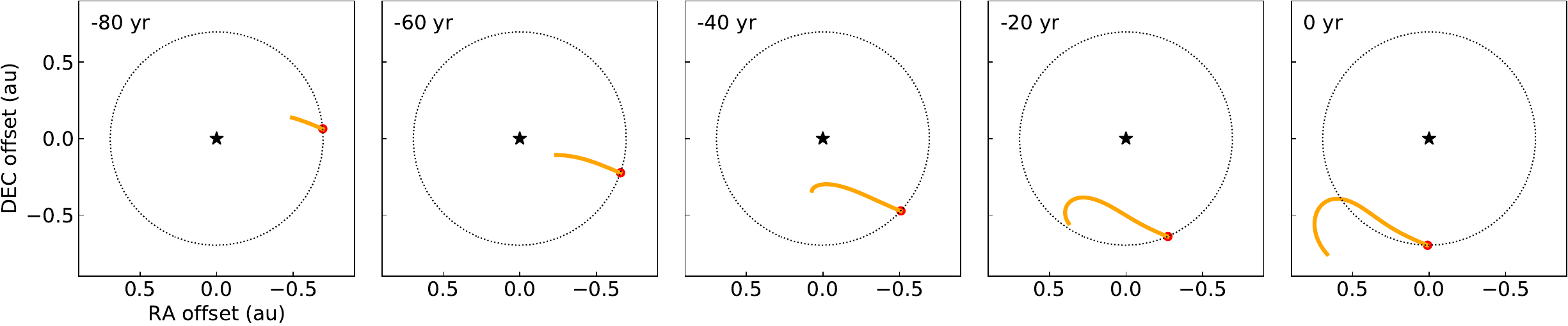}
\caption{Time evolution of the trajectories from 80 yr ago to present time. The black dotted line and star mean the 42 au orbit and central star. The red point indicates the location of the protoplanet at that time.   }
\label{fig:trajects}
\end{figure*}
The outflow trajectory from 80 yr ago until now is plotted in Figure \ref{fig:trajects}. Unlike protostellar outflows, the morphology of planet-driven outflows changes time by time because of the orbital motion of the driving source.


\bibliography{sample631}{}
\bibliographystyle{aasjournal}



\end{document}